\documentclass[conference]{IEEEtran}

\usepackage{booktabs}
\usepackage{xcolor}
\usepackage[utf8]{inputenc}

\usepackage[numbers,sort&compress,nonamebreak]{natbib}
\usepackage{microtype}
\usepackage{hyperref}
\usepackage{balance}
\usepackage{url}

\usepackage[T1]{fontenc} %
\usepackage[cmex10]{amsmath}
\usepackage{amssymb}
\usepackage[cmintegrals]{newtxmath}
\usepackage{bm} %

\usepackage{graphicx}
\usepackage{url}
\usepackage{supertabular}
\usepackage{color}

\renewcommand\vec{\mathbf}
\newcommand{\cd}[1]{\texttt{#1}}

\IEEEoverridecommandlockouts%
\IEEEpubid{\makebox[\columnwidth]{\scriptsize\color{red}\shortstack{%
    Author post-print. Paper published in the proceedings of the 30th International\\%
    Symposium on Computer Architecture and High Performance Computing.\\%
    \textcopyright{} 2018 IEEE - DOI:\href{https://doi.org/10.1109/CAHPC.2018.8645881}{10.1109/CAHPC.2018.8645881}\\\vspace*{-3em}%
}} \hspace{\columnsep}\makebox[\columnwidth]{ }}

\title{An argument in favor of strong scaling for deep neural networks with small datasets}
\author{
    \IEEEauthorblockN{Renato L. de F. Cunha, Eduardo R. Rodrigues, Matheus Palhares Viana, Dario Augusto Borges Oliveira}
    \IEEEauthorblockA{IBM Research}
}

\begin{document}
\maketitle

\begin{abstract}

In recent years, with the popularization of deep learning frameworks and large datasets,
researchers have started parallelizing their models in order to train faster.
This is crucially important, because they typically explore many hyperparameters in order
to find the best ones for their applications. This process is time consuming
and, consequently, speeding up training improves productivity.  One approach to
parallelize deep learning models followed by many researchers is based on weak scaling.
The minibatches increase in size as new GPUs are added to the system. In addition, new
learning rates schedules have been proposed to fix optimization issues that occur
with large minibatch sizes. In this paper, however, we show that the recommendations provided
by recent work do not apply to models that lack large datasets. In fact, we
argument in favor of using strong scaling for achieving reliable performance in
such cases. We evaluated our approach with up to 32 GPUs and show that weak
scaling not only does not have the same accuracy as the sequential model, it also
fails to converge most of time. Meanwhile, strong scaling has good scalability
while \emph{having exactly the same accuracy of a sequential implementation}.

\end{abstract}


\section{Introduction}\label{sec:introduction}

Neural networks have gained popularity in recent years~\cite{NIPS2012_4824,lecun2015deep}. Several models have been
created to perform diverse tasks, such as image classification~\cite{ren2015faster}, segmentation~\cite{danet},
language translation~\cite{bahdanau2014neural}, and even playing games~\cite{DBLP:journals/corr/MnihKSGAWR13,mnih2015human}. Two factors have been presented as the
reasons for this new increase in neural network interest: (1) large datasets and (2) computational power growth.

Large datasets allow for the adjustments of the large number of parameters that deep neural networks
have. Moreover, the large computational power that recent CPUs and accelerators provide, enables training of
these networks in a reasonable amount of time. Still, parallel techniques have been devised to accelerate further the
training of single large neural networks that uses several processing units at the same time.

Many techniques have been proposed in the literature to optimize the performance of parallel neural network pipelines,
as datasets increase in size. One popular technique is to replicate the neural network across several processing
units, divide the training set among these units and update gradients synchronously~\cite{DBLP:journals/corr/abs-1708-02188} or asynchronously~\cite{chilimbi2014project}. 
Since the datasets are typically very large, one common approach is to increase the minibatch size as the number of 
processing units increases. At each step, the minibatch is divided among the processing units and each of them
computes the stochastic gradient descent updates of its part of the minibatch. After that, the gradients are averaged and
distributed back to the processing units, so that they can update the weights.

The increase of minibatch size as the number of processing units increases can be viewed as weak scaling~\cite{foster1995designing}, that is,
instead of reducing execution time as the number or processors/accelerators goes up, one tries to increase the amount
of data processed as the number of processing units increases while keeping the execution time constant. This is an approach
known in High Performance Computing (HPC) as Gustafson's law~\cite{gustafson1988reevaluating}. Still, some researchers have shown~\cite{goyal2017accurate} that
one needs to adjust the learning rate (and other parameters) to keep the accuracy quality.

Despite all the progress in parallelizing deep neural network models, little has been done to speed up models for which
there is little data. One may argue that in an age of \emph{Big Data}, this problem does not exist, since data is
abundant. In addition, even if that is not the case, models with little data are fast to execute and do not need to be
parallelized. However, in this paper we show an application for which the training set is small, but the training time is
long. Moreover, we propose a strategy to run the model in parallel that goes against the wisdom proposed by all other
strategies, that is, the efficiency is best with a strong scaling approach rather than a weak one.

In this paper we present our parallel execution strategy and the results with a medical imaging application, which we also describe.
However, we expect that similar results can be obtained in other applications that also have small training
datasets. This paper is divided as follows: in the next section we present a short background on why we think the advice to
run parallel deep learning models is not complete. Section~\ref{sec:related-work} reviews the literature on parallel execution of
deep learning models. Our methodology and the alternatives we compare it against are shown in Section~\ref{sec:methodology}.
Section~\ref{sec:app} presents the application used in our experiments. This application has a small training set, but takes
very long to execute sequentially. Our experiments are presented in Section~\ref{sec:experiments} and its results are discussed
in the section after that. The final section presents our final remarks and conclusions.

\section{Background}\label{sec:background}%

Neural Network models are appealing due to their ability to approximate
arbitrary functions~\cite{hornik1991approximation}. The advent of higher-level frameworks
for the definition and training of Neural Networks (such as TensorFlow, PyTorch,
and CNTK), together with the increase of computational power have led to a raised
interest in Neural Networks and faster execution of ever-growing models.
Although there have been early attempts at parallelizing Neural Networks, they
were usually tied to the architecture and implementation of specific neural
networks. For example, the original AlexNet \cite{NIPS2012_4824} would run in
two GPUs, because of the lack of single GPU memory to hold the data and parameters
needed to train the model. The network architecture was partitioned between the
GPUs, rather than being duplicated in them, as it has become popular now.

One of the enabling technologies to the recent boom of popularity for Neural
Networks is Big Data. Hence, benchmarks and applications are set up in
ways that the networks can exploit the availability of such data. For example,
ImageNet has 14,197,122 examples divided into 21,841 categories. Recent work in the
literature has devised rules for setting up experiments so that parallelization
has a faster rate of convergence. However, such rules might only apply to models
that have access to large amounts of data, since they tend to rely on increasing
the minibatch size\footnote{The number of samples processed by all workers in each step of the
optimization procedure.} with the increase in the number of workers. This makes sense
in a data rich environment such as computer vision, in which labeled pictures and videos
have become plentiful.

However, many applications rely on hard to obtain samples, in which 
the associated cost to obtain samples is still very high. This is particularly true
for medical imaging applications \cite{shen2017deep}. Nonetheless, the training time can still be very long, despite 
the lack of large training sets. Speeding up these applications will allow practitioners to explore
a wider range of hyperparameters and, possibly, to find better models. In this paper, we
investigate how the parallelization rules found in the literature impact models that do not have
access to large amounts of data, and contrast them with a strong scaling
approach, in which the number of workers increases, the minibatch size is held
constant.
\section{Related Work}\label{sec:related-work}

Goyal \textit{et al.}~\cite{goyal2017accurate} present a strategy to train ResNet-50~\cite{DBLP:journals/corr/HeZRS15} in one hour using
the ImageNet dataset. They employed a distributed synchronous Stochastic Gradient Descent (SGD) approach with up to 256 GPUs. They
used a large minibatch---8192 examples---and reached an accuracy similar to a much smaller minibatch of 256 examples.

In order to achieve this result, Goyal \textit{et al.} proposed a linear scaling rule for the learning rate and a warmup scheme. That was
because they found that optimization difficulties were the major issue with the large minibatches rather than poor generalizations, as 
previous results had suggested. Moreover, the large minibatches allowed them to increase the number of GPUs in a way that the number of
examples processed increased linearly with the number of GPUs. 

If one considers the problem size as the minibatch, then this strategy is similar to the weak scaling approach to scalability, in which one keeps
the problem size per processor constant as the number of processors increases. A different strategy would be to increase the number of
processors (or GPUs) and keep the minibatch size constant. This is guaranteed to produce the same accuracy as the sequential execution, but
may result in saturated speedups 

Cho \textit{et al.}~\cite{DBLP:journals/corr/abs-1708-02188} describe a topology aware distributed synchronous gradient descent strategy. In
it, the communication of gradients is scheduled so that it maximizes the use of the available communication channels. Most of the
communication in a distributed SGD optimization process occurs in the first stages of the gradient communication. As the gradient communication progresses, 
fewer messages are sent. Consequently, one can use this fact to schedule the communication to use the fastest channels at the beginning and
progressively use the slower channels.

Tensorflow has also parallel execution capabilities built-in~\cite{abadi2016tensorflow}. In this framework, the programmer specifies a set
of operations to be placed in the available devices. These devices can be processors and accelerators (GPUs and TPUs) that are distributed
across a cluster of computers. A common strategy is to split the data and processing units into parameter servers and workers nodes. 
The parameter servers are responsible for holding the network weights and other parameters, while the workers are responsible for computing
the forward and backward passes of the model.

The Tensorflow framework has grown in great popularity. However, its parallelization capabilities have been criticized for its unnecessary 
complexity. One alternative was proposed by Sergeev and Balso~\cite{sergeev2018horovod}. They used a simplified strategy based on MPI to 
parallelize and run distributed SGD models. A key feature of the approach is to combine small messages so that it better uses the
network.

In medical imaging, the current state-of-art for classification and segmentation of 3D exams are 3D deep learning models \cite{shen2017deep}.
However, the widely known computational burden of such models due to 3D convolutions, in many times hinders the development of real-time
systems for aiding in the diagnosis of 3D exams. This creates opportunities for methods that aim at providing time-efficient support to process
3D models in parallel.
\section{Methodology}\label{sec:methodology}

In this section, we describe our approach to parallelize the training procedure and the alternatives found in
the literature, which we compare ours with. We start with a general description of the Stochastic Gradient Descent (SGD) method, and the
two possible ways to deal with the minibatch sizes as one increases the number of processing units. In
addition, we review the procedures for adjusting the learning rates as one increases the number of GPUs and
a warmup strategy. Finally, we show the evaluation strategy used to compare the different alternatives.

The SGD update has the form
\begin{equation}
    w_{t+1} = w_t - \eta \frac{1}{|\mathcal{B}|}\sum_{x \in \mathcal{B}}\nabla l(x,w_t)\text{,}
\label{eq:seq}
\end{equation}
where $w$ are the weights, $\eta$ is the learning rate, $\mathcal{B}$ is the minibatch, and $l$ is the
loss of each example $x$. 

Since the gradient of the loss $l$ is independent for each example, the weight updates can be computed
in a bulk-synchronous manner. The initial weights ($w_0$) are set to be identical in all $N$ processing units. Then,
the minibatch ($\mathcal{B}'$) is divided into $N$ sets ($B_i$s) distributed across the processing units, such that:

$$
\bigcup_{i=0}^{N-1} B_i = \mathcal{B}'
$$
$$
\bigcap_{i=0}^{N-1} B_i = \varnothing
$$

The $B_i$s are typically of the same size. Then, the SGD updates can be trivially rewritten as:

\begin{equation}
w_{t+1} = w_t - \eta \frac{1}{|\mathcal{B}'|}\sum_{i=0}^{N-1}\Big(\sum_{x \in B_i}\nabla l(x,w_t) \Big)
\label{eq:par}
\end{equation}

\noindent
in which the summation inside the parenthesis is computed in parallel across the processing units and
its average is performed in an \emph{all reduce} operation. \emph{Reduce} is a communication
primitive that reduces a set of values (or vectors) distributed across some processing units to a single value (or single vector). For
example, average and maximum (or element-wise average and maximum) are possible reduction operations. In particular, distributed SGD communicates
the averages and distributes them back to all processing units, in what is called an \emph{all} reduce
operation. All processors, then, update their copies of the weights.

Our parallelization strategy keeps the size of the minibatch $\mathcal{B}'$ the same as the number of
processing units varies, i.e. $\mathcal{B}' = \mathcal{B}$. This makes the updates in Equation~\eqref{eq:par} the same as in Equation~\eqref{eq:seq}. 
Indeed, this guarantees that the optimization follows the same path through the loss landscape irrespective to
the number of processing units used. However, the ratio of communication to computation increases as one 
increases the number of processing units, and this hurts the speedup:

$$
\mathrm{Speedup}(n) = \frac{T(1)}{T_S + T_P/n + c\,\log(n)} = \mathcal{O}\Big(\frac{1}{\log(n)}\Big)
$$

\noindent
where $n$ is the number of processing units, $T(1)$ is the sequential execution time, $T_S$ is the time that parts of the parallel
implementation still run sequentially (e.g. weight initialization), and $T_P$ is the time spent in parallel regions, but without communication. 
Since the communication is at best logarithmic,
the speedup will saturate for a large enough number of processing units. In order to avoid this issue, most 
researchers have used large minibatches \cite{goyal2017accurate,DBLP:journals/corr/abs-1708-02188}. 

Deep learning practitioners have been advised to increase batch sizes not only to improve parallelism
\cite{goyal2017accurate} but also to reduce the number of parameter updates while
achieving the same accuracy \cite{smith2017don}. Still, in order to maintain accuracy and the same workload as one increases the
number of processing units, the learning rate must be adjusted. Goyal \textit{et al.} \cite{goyal2017accurate}
propose a \emph{linear scaling rule} that is used by others \cite{DBLP:journals/corr/abs-1708-02188,DBLP:journals/corr/abs-1708-03888}.
This rule instructs the user to increase the learning rate by $k$ whenever the minibatch size is increase by that same $k$.
The intuitive reason to justify this rule is that taking $k$ steps with learning rate $\eta$ and minibatch size $|B_j|$:
\begin{equation}
w_{t+k} = w_t - \eta \frac{1}{|B_j|}\sum_{j<k}\sum_{x \in B_j}\nabla l(x,w_{t+j})
\end{equation}
is equivalent to taking a single step with learning rate $\eta$ times $k$ and minibatch of size $|\mathcal{B}|$, where $\mathcal{B} = \cup_j B_j$:
\begin{equation}
w_{t+1} = w_t - k \eta \frac{1}{|\mathcal{B}|}\sum_{j<k}\sum_{x \in B_j}\nabla l(x,w_{t})\text{.}
\label{eq:pareq}
\end{equation}

As pointed out by Goyal \textit{et al.}~\cite{goyal2017accurate}, it is clear that these two updates will probably not be identical. However,
the authors argue that the only case in which they can actually be equal is precisely when the learning rate in Equation~\eqref{eq:pareq} is
$k \eta$. Moreover, the authors show a limit of the minibatch size that suggests a point beyond which the differences between
the two updates are too large to yield similar results.

In addition to the linear scaling rule, a warmup procedure is also typically used as the minibatch sizes increase. This procedure
increases the learning rate gradually in the beginning of training until it reaches the value $k \eta$. From that point on, the
learning rate is kept constant at $k \eta$. This is needed because the network weights change rapidly in early stages of training.
Consequently, large learning rates at that stage may cause optimization problems, as noted by Goyal \textit{et al.}~\cite{goyal2017accurate}.

Neither the linear scaling rule nor the warmup procedure is needed in our
strategy, since the learning rate is the same as in the sequential execution.
Moreover, since we keep the effective minibatch size constant, we know our
strategy will follow the same optimization path in the error surface
a sequential model would follow. We will compare our parallelization strategy
with one that uses large minibatches and the linear scaling rule, with and
without the warmup procedure. In addition, we compare with another approach that
uses large minibatches but keeps using the same learning rate as the sequential
execution. This will be done in the context of an application that has a small
training set, but that still takes very long to train. We present this
application in the next section.

In order to compare the proposed approach with the alternatives, we evaluated
the time to loss. We trained the model described in the next section with
different number of GPUs until the model reached a certain target loss.
Comparing the time to execute a given number of epochs would be incorrect in
this context. That is because the weak scaling approach will definitely be
faster when using this criterion, since each evaluation will consume much larger
blocks of the batch. However, they will not achieve the same loss as the strong
scaling approach.
\section{Application}\label{sec:app}

\begin{figure*}[!t]
    \centering
    \includegraphics[width=1.0\linewidth]{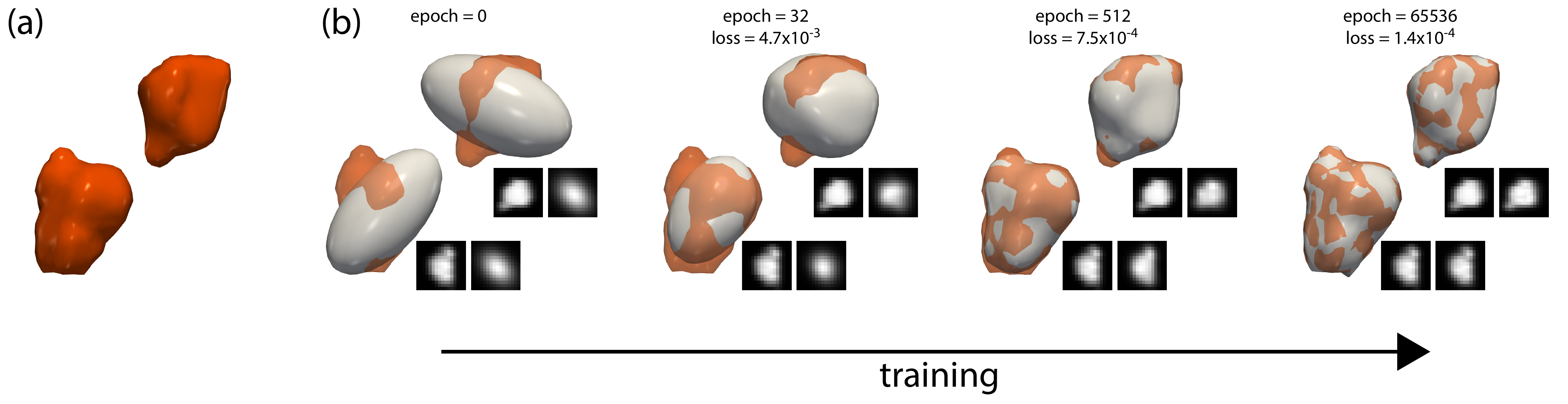}
    \caption{
        \textbf(a) Two examples of 3D lung nodules from the LUNA dataset.
        \textbf(b) Autoencoder estimation of the two lung nodules shown in (a)
        modeled as a mixture of \textit{trivariate normal distributions}. The
        insets correspond to the central slice of the original 3D nodule image
        on the left and the central slice of the estimate density on the right.
    }
    \label{fig:prediction}
\end{figure*}

As an application, we chose a dataset of 3D lung nodules obtained from the LUNA
challenge\footnote{\texttt{\url{https://luna.grand-challenge.org/}}}. The
dataset consists of a 888 Computed Tomography scans for which experts have annotated the 3D
coordinates of 2100 nodules greater than 3 mm. Nodules are grouped in
four different categories depending on their visual appearance, which is usually
correlated with their malignancy. The main goal of researchers working on this
dataset is to develop a system for automated diagnosis as accurate as
possible.  However, the small number of nodules available in each category
prevents us from training 3D image classifiers that achieve high-precision
diagnosis.

To overcome the limitation of small data, a common strategy is to perform data
augmentation to increase the number of images in our training set. To go beyond
the traditional translational, rotational and color
transformations~\cite{NIPS2012_4824}, we decided to model the lung nodules as 3D
Gaussian mixture models. Upon having learnt the typical parameters that describe
different types of lung nodules, we could generate as many nodules as required
to achieve the required diagnosis accuracy.

Mixture models are used to describe systems composed by subpopulations within an
overall population. Gaussian mixture models in particular, are used in different
applications, ranging from to speech recognition~\cite{reynolds2000speaker}, electron and atomic
position~\cite{kawabata2008multiple}, cell biology~\cite{khan2012gamma}, and others. The $n$-dimensional
multivariate Gaussian distributions is written as
\begin{equation}
    {\cal N}(\vec{x},\vec{\mu},\mathbf{\Sigma}) = \frac{1}{\sqrt{2\pi\mathbf{\Sigma}}}
    \exp\left[-\frac{1}{2}\left(\vec{x}-\vec{\mu}\right)^T\mathbf{\Sigma}^{-1}\left(\vec{x}-\vec{\mu}\right)\right]\text{,}
    \label{eq:multidensity}
\end{equation}
where $\vec{\mu}$ is the mean and $\mathbf{\Sigma}$ is the positive-definite covariance matrix
\begin{equation}
    \mathbf{\Sigma} =
    \begin{bmatrix} 
        \sigma_1^2 & \rho_{1,2}\sigma_1\sigma_2 & \cdots & \rho_{1,n}\sigma_1\sigma_n \\
        \vdots & \vdots &  \ddots & \vdots \\
        \rho_{n,1}\sigma_n\sigma_1 & \rho_{n,2}\sigma_n\sigma_2 & \cdots       & \sigma_n^2
    \end{bmatrix}.
    \label{eq:covm}
\end{equation}

The mixture of $K$ multivariate Gaussian distributions can be expressed as
\begin{equation}
    {\cal M}(\vec{x}) = \sum_{i=1}^K \alpha_i\; {\cal N}(\vec{x},\vec{\mu}_i,\mathbf{\Sigma}_i),
    \label{eq:mixmultidensity}
\end{equation}
such that
\begin{equation}
    \sum_{i=1}^K\alpha_i = 1\text{.}
    \label{eq:constraint}
\end{equation}

We use an autoencoder architecture for end-to-end unsupervised learning. The parameters of mixture of multivariate Gaussian distributions are learnt from the encoding generated by the autoencoder to represent the 3D input data. At the same time, the parameters are used to reconstruct the input using Equation~\eqref{eq:multidensity}. The latent encoding $\vec{y}$ is generated by a convolutional encoder for $n-$dimensional input data $\vec{x}$, where $n=1$, $2$ or $3$, according to
\begin{equation}
    \vec{y} = \Phi\left(\mathbf{W}_0\vec{x}+\vec{b}_0\right)\text{.}
\end{equation}

The latent representation $\vec{y}$, which is of size ${\ell}\times 1$, is then used as input for dense layers that in parallel estimate the parameters $\alpha_i$, $\vec{\mu}_i$ and $\mathbf{\Sigma}_i$, $0\leq i \leq K$. Because of the constraint expressed as Equation~\eqref{eq:constraint}, we use \cd{softmax} activation for estimating $\alpha's$. We used \cd{tanh} for estimating the components of the mean parameters $\vec{\mu}_i$ which lie in the range $[-1,1]$ representing the domain of the input data. Standard deviation and correlation parameters are estimated through \cd{sigmoid} activation and they are composed to build the covariance matrices given by Equation~\ref{eq:covm}. Therefore, the parameters of our mixture model can be expressed as

\begin{eqnarray}
    \vec{\alpha} & = & \Theta\left(\mathbf{W}_1\vec{y}+\vec{b}_1\right),\\
    \vec{\mu}_1,\dots,\vec{\mu}_K & = & \Xi\left(\mathbf{W}_2\vec{y}+\vec{b}_2\right),\\
    \vec{\sigma}_1,\dots,\vec{\sigma}_K & = & \Psi\left(\mathbf{W}_3\vec{y}+\vec{b}_3\right),\\
    \vec{\rho}_{1,2},\dots,\vec{\rho}_{K-1,K} & = & \Psi\left(\mathbf{W}_4\vec{y}+\vec{b}_4\right).
\end{eqnarray}

In total, our architecture estimates $1/2(n^2+3n+2)K$ parameters from the latent representation of 3D lung nodules used as input. The last layer in our architecture, called \textit{DeepDensity}, uses the estimated parameters to create the $n$-dimensional density map in Equation~\eqref{eq:mixmultidensity}. The resulting density map is compared to the input image according to a loss function of our choice.

As the training progresses, the autoencoder rapidly learns the overall shapes of
3D nodules. However, the very fine details, crucial for creating realistic
synthetic nodules during data augmentation, are learnt in a much slower time
scale. Figure~\ref{fig:prediction} shows an example of the learning process.
This requires a very long training time and makes the pipeline prohibitive in
practice. For this reason, we believe that this application is a good example of
how parallelization can not only speed-up training processes, but make them
feasible.
\section{Experiments}\label{sec:experiments}

Recall the application under investigation in this paper is an unsupervised
one: its objective is to learn a Gaussian Mixture Model that approximates the
distribution of the data.  Hence, traditionally used metrics cannot be used to
evaluate the model's performance. Due to that, our experiments measure the
time it takes to reach a given loss. Traditionally, other researchers used the
number of epochs (full passes over the entire data set) as a measure of time, but
since there can be differences in the way data is split, we decided to use the
\emph{time in seconds} to reach a given loss.

Our experiments were performed in a cluster of POWER8 machines running Red Hat
Enterprise Linux release 7.4. POWER8 machines support simultaneous
multi-threading (SMT) values in the set $[1, 2, 4, 8]$. All hosts in which we
performed experiments had SMT set to $8$. Therefore, although hosts have 24
cores, $24 \times 8 = 192$ logical cores are available to applications. Such SMT
values were set by the administrators of the cluster to increase cluster
utilization.  Each host also has four NVIDIA Tesla K80 GPUs. To facilitate the
scheduling of our jobs, we allocated at most two GPUs per execution node, with
the exception of executions with a single GPU\@. For communication between
hosts, we used OpenMPI version 2.1.3, with ranks in the same hosts communicating
with shared memory, and ranks in different hosts communicating over TCP/IP using
Gigabit Ethernet. The application under investigation was written and evaluated
using Keras 2.1.6 and TensorFlow 1.7.1.

The target (validation) loss for these experiments was $0.0016384$, which
roughly corresponds to a hundred epochs when running the sequential version of
the code. The validation set is composed of 256 samples, and the rest of the
dataset is used for training.  All experiments had a maximum limit of 400 passes
over the training set (epochs).  Hence, if a model failed to reach the target
loss in 400 epochs, its execution was stopped. We also stopped execution in
case any infinite or \textsc{NaN} (Not a Number) value was found when computing
losses. At the end of each epoch, we logged the training loss, the validation
loss, and the time spent since the beginning of training.

For training the neural network, we used the Adam~\cite{kingma2014adam}
optimizer with base learning rate $\eta = 0.00105$, and other hyperparameters
set to default values ($\beta_1=0.9$, $\beta_2=0.999$,
$\epsilon=1\times10^{-8}$, and decay = $0$).  This learning rate was kept the
same for the strong scaling and weak scaling experiments. For the learning
scaling rule and warmup experiments, the learning rate was set to $k\eta$, where
$k$ is the number of GPUs used, with the warmup experiments increasing the
learning rate linearly up to $k\eta$ over five epochs, as described in the
literature~\cite{goyal2017accurate}. Notice that when only one GPU is used, all
scaling strategies behave the same, since $k\eta=\eta$. For the application
domain, we used cubes of size 16, $K=50$ distributions, \cd{mean square log
error} as loss function, and three convolutional layers with 32 filters of size
$7\times7\times7$, $5\times5\times5$, and $3\times3\times3$ for each layer
respectively.

An interesting aspect of the chosen learning rate is that further increasing it
in the sequential model causes divergence, which is aligned to the guideline
given in the literature of using the largest learning rate that still makes
the optimization converge~\cite{goh2017why}. Therefore, it is expected that
increasing the learning rate as in the literature should cause the model to
diverge.

\section{Results}\label{sec:results}

\begin{figure}
    \centering
    \includegraphics[width=\linewidth]{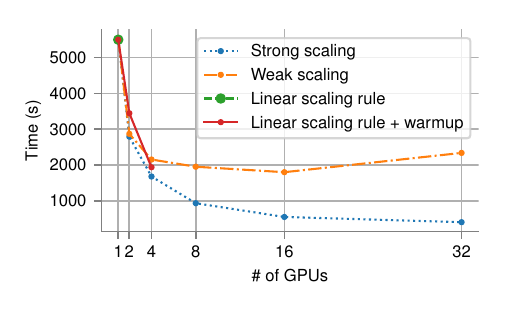}
    \caption{%
        Time to reach target accuracy for the various scaling modes. We
        evaluated the time to reach the target loss for $1, 2, 4, 8, 16$, and
        $32$ GPUs. In the strong scaling test, we maintained the batch size
        constant over all executions. Therefore, with one GPU, the GPU used a
        batch of size 32. With two GPUs, each GPU used a batch size of 16, and
        so on. For the weak scaling tests (weak scaling, linear scaling rule, and
        linear scaling rule + warmup), the total batch size increased with each
        GPU and, for two GPUs, the batch size was 64; for three, it was 96, and
        so on. The linear scaling rule line is not visible in the plot because
        it diverged in \emph{all} experiments with a number of GPUs different
        from 1.
    }\label{fig:times}
\end{figure}

Figure~\ref{fig:times} shows the results of the experiments described in the
previous section. In the figure, it can be seen that only the experiments that
maintained a constant learning rate (strong scaling and weak scaling) converged
to a solution in all GPU configurations. The warmup strategy only converged with
one, two, and four GPUs, whereas the linear scaling rule strategy failed to
converge in \emph{all} GPU configurations but one, suggesting that it is a poor
scaling strategy when a well-tuned learning rate is used. The reason for
divergence is that the learning rate was set too high. With the warmup
experiment, it appears warmup did indeed help convergence with four GPUs, as
convergence time was smaller than that of the weak scaling strategy. Still, no
scaling strategy was faster than the strong scaling one, suggesting that
although warmup can help convergence times, for our application, using strong
scaling yields faster and more reliable results. In the weak scaling case, we
see an increase in times for 32 GPUs. We attribute such an increase in
processing times to the noise introduced when using bigger batch sizes.

\begin{figure}
    \centering
    \includegraphics[width=\linewidth]{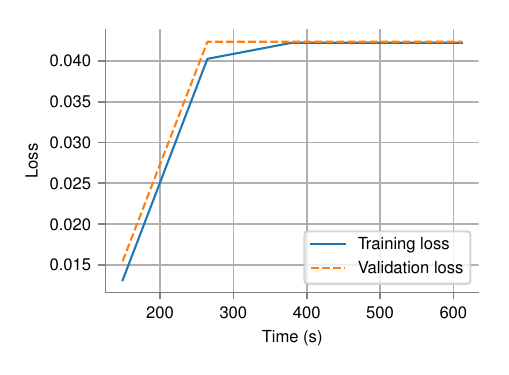}
    \caption{%
        Behavior of the loss with warmup when warming up for five epochs and
        using eight GPUs. As can be seen in the graph, as the learning rate
        increases due to warmup, the loss also increases with it, until hitting
        a ceiling of approximately $0.042$. After hitting this ceiling, the loss
        stays mostly the same until the alloted time is up for this task.
    }\label{fig:warmup-8gpu}
\end{figure}

A deeper analysis in the warmup case is necessary, since it has components of
both convergence and divergence. When we analyze the training and validation
losses computed by the model in the case with 8 GPUs, we see an interesting
pattern: as the learning rate increases, the loss increases as well, reaching
a ceiling of around $0.042$ and staying there for the rest of the experiment.
This effect is shown in Figure~\ref{fig:warmup-8gpu}.  Whereas in the 16 GPU
case, neither loss ever changes, staying at the $0.042$ level as well. For the
32 GPU case, the losses diverge to $+\infty$ within the first epoch. We believe
such behavior is, again, related to the increase in the learning rates, causing
the optimization process to oscillate and, ultimately, diverge. If the number
of epochs during which warmup is active is increased to 10, the same phenomenon
happens, with the experiment with 8 GPUs and subsequent ones failing to converge.

\begin{figure}
    \centering
    \includegraphics[width=\linewidth]{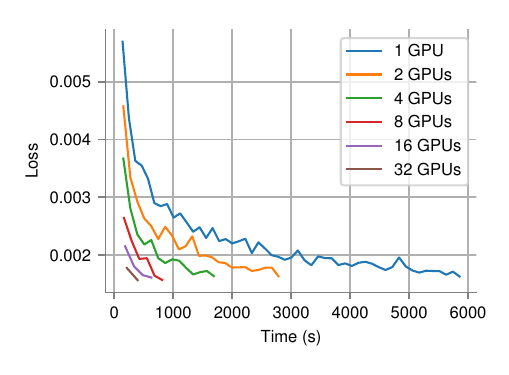}
    \caption{%
        Behavior of the loss of the model when strong scaling is used. The figure shows
        the losses for 1, 2, 4, 8, 16, and 32 GPUs. It can be seen in that the loss
        tends to have a sharp decrease in the beginning followed by a slower decay in
        later iterations. This effect is not pronounced in the 32 GPUs case because
        the model converges in just two iterations in this case.
    }\label{fig:losses-strong}
\end{figure}

\begin{figure}
    \centering
    \includegraphics[width=\linewidth]{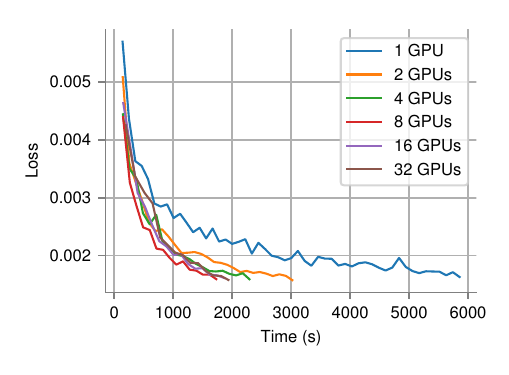}
    \caption{%
        Behavior of the loss of the model when weak scaling without learning rate
        scaling is used. The figure shows the losses for 1, 2, 4, 8, 16, and 32
        GPUs. The losses overlap each other, suggesting there is not much to
        gain from such a scaling strategy, since losses do not improve much.
    }\label{fig:losses-weak}
\end{figure}

Both in the strong scaling and in the sequential cases, the loss profile is such
that there is a sharp decrease in the beginning, followed by a slow, but steady,
decrease of the loss. This corresponds to the model rapidly learning the overall
shapes of the nodules, followed by a slow learning process of the fine details,
as described at the end of Section~\ref{sec:app}. The losses of such case are
shown in Figure~\ref{fig:losses-strong}. Since the only weak scaling strategy
that converges in all cases is the one \emph{without} learning rate scaling,
we show its loss profile in Figure~\ref{fig:losses-weak}. As can be seen in the
figure, increasing parallelism does not help much, and the only significant
improvement comes from increasing parallelism from one to two.

\begin{table}
\centering
\caption{%
    95\% confidence intervals ($n = 11$) for strong scaling efficiency of the
    various GPU configurations.
}\label{tab:strong-efficiency}
\begin{tabular}{lr}
    \toprule
    GPUs &  Efficiency (\%) \\ %
    \midrule
       1 &  100.00 $\pm$  0    \\ %
       2 &   95.03 $\pm$ 10.19 \\ %
       4 &   84.90 $\pm$  9.53 \\ %
       8 &   75.06 $\pm$  8.13 \\ %
      16 &   59.92 $\pm$ 10.04 \\ %
      32 &   44.05 $\pm$  4.86 \\ %
    \bottomrule
\end{tabular}
\end{table}

With regards to scaling efficiency, we noticed that in some cases one can see
a super-linear speedup when transitioning from one GPU to two GPUs, caused by
the stochasticity of the evaluation process.  Table~\ref{tab:strong-efficiency}
shows the efficiency for the various GPU configurations for the strong scaling
experiments. We computed efficiency as
\[
\mathrm{efficiency} = \frac{T_1}{n T_n}100\% \text{,}
\]
where $T_1$ is the time to process the data with one GPU, $n$ is the number of
GPUs, and $T_n$ is the time to process the data with $n$ GPUs. In the table, we
show 95\% confidence intervals of the average efficiency of eleven trials.
When we observe the scaling efficiency of the other parallelization strategies,
we see that they, as expected, are less efficient than the strong scaling
parallelization strategy. Again, the strong scaling strategy does not suffer
from divergence issues. The results are summarized in
Table~\ref{tab:weak-scaling}.  In the table, apart from the number of GPUs, the
numbers represent scaling efficiency in \%.

\begin{table}
    \centering
    \caption{%
        Efficiency numbers for the weak scaling experiments. All numbers are
        smaller than their strong scaling counterparts. Entries with NaN correspond
        to experiments in which the neural network failed to converge to the target
        loss of $0.0016384$.
    }\label{tab:weak-scaling}
    \begin{tabular}{lrrrr}
    \toprule
      GPUs &  Weak scaling &  Warmup &  Linear scaling rule \\
    \midrule
         1 &    100.00     &  100.00 &  100.0 \\
         2 &     96.96     &   84.95 &    NaN \\
         4 &     63.86     &   75.49 &    NaN \\
         8 &     42.32     &     NaN &    NaN \\
        16 &     19.65     &     NaN &    NaN \\
        32 &      9.45     &     NaN &    NaN \\
    \bottomrule
    \end{tabular}
\end{table}

\section{Conclusion}\label{sec:conclusion}

In this paper we tackled the problem of scaling the training of an application
that lacks big data for training. As a consequence, we have argued and shown
that, at least for this particular application, recently proposed scaling
strategies in the literature fail to converge in most cases. We have also shown
that, apart from its counterintuitive aspect, given the recent suggestions from
the literature, strong scaling (in which the size of the minibatch is held
constant) is the best strategy for this application.

We have shown that even modest models with small data can demand large amounts
of computing, warranting a parallel implementation not only for the model to
train faster, but also to make it a viable solution with more chances of adoption
in production environments.

We believe some results reported in the literature may not transfer to problems
that lack large amounts of data, and may be biased towards the ImageNet
benchmark. As we have seen, the guidelines recently found in the literature fail
for at least one application with small data. Although we tested with a single
application, we expect the results presented in this paper to generalize to
applications \emph{with small data}.

Our main result in this paper was to show that a strong scaling approach is both
faster than weak scaling alternatives, and that it reliably converges to
a solution, which can not be said about the tested weak scaling approaches.
 
\bibliographystyle{IEEEtran}
\balance
\bibliography{references}

\end{document}